\documentclass{mn2e}
\usepackage{epsfig}

\newcommand{\ltsimeq}{\raisebox{-0.6ex}{$\,\stackrel
        {\raisebox{-.2ex}{$\textstyle <$}}{\sim}\,$}}
\newcommand{\gtsimeq}{\raisebox{-0.6ex}{$\,\stackrel
        {\raisebox{-.2ex}{$\textstyle >$}}{\sim}\,$}}

\def\le{{L_{\rm Edd}}}
\def\msun{{\rm M_{\odot}}}

\def\ro{{R_{\rm outflow}}}
\def\rs{{R_{\rm s}}}
\def\mo{{\dot M_{\rm out}}}
\def\me{{\dot M_{\rm Edd}}}
\title[The Nature of SS433]
{The Nature of SS433 and the Ultraluminous X--ray Sources}
\author[M.C. Begelman, A.R. King \& J.E. Pringle]
{M. C. Begelman$^{1,3}$, A.R. King$^2$ \& J.E. Pringle$^{2,3}$ 
\\
$^1$JILA, University of Colorado, Boulder, CO 80309-0440, USA\\
$^2$Department of Physics and Astronomy, University of Leicester,
Leicester, LE1~7RH, UK\\
$^3$Institute of Astronomy, Madingley Road, Cambridge, CB3 0HA, UK\\
}

\begin{document}

\maketitle
\begin{abstract}


The periodic precession (162--day) and nodding (6.3--day) motions of
the jets in SS433 are driven in the outer regions of the disc, whereas
the jets themselves, being relativistic, are launched near the black
hole at the disc centre. Given that the nutation period is comparable
to the dynamical timescale in the outer regions of the disc, it seems
unlikely that these periods can be communicated efficiently to the
disc centre. Here we propose that the massive outflow observed in
SS433 is launched at large radii in the disc, about 1/10 of the outer
disc edge, and that it is this outflow which responds to the
oscillations of the outer disc and determines the direction of the
jets. The massive outflow is launched at large radius because the mass
transfer rate is hyper-Eddington. This implies not only that the total
luminosity of SS433 exceeds $\le$ by a considerable factor, but also
that the radiative output is collimated along the outflow. We thus suggest
that SS433 is an ultraluminous X--ray source (ULX) viewed `from the
side'. We also suggest that the obscured {\it INTEGRAL} sources may be
SS433--like objects, but with slightly lower mass transfer rates.

\end{abstract}


\section{Introduction}

The extraordinary properties of SS433 have long attracted
attention. This 13.1--day binary system shows a pair of
well--collimated jets with velocity $\pm 0.26c$ which precess on a
cone of half--angle $\theta \simeq 20^{\circ}$ with period $P_{\rm pr}
= 162$~d (Margon, 1979, 1984; Abell \& Margon, 1979; Milgrom, 1979;
Fabian \& Rees, 1979; see Fabrika, 2006 for a recent review of
observational properties). In addition to the jets there is a powerful
outflow with velocity $v_w \sim 1500$~km\, s$^{-1}$ which may have
inflated the surrounding nebula W50 (Begelman et al., 1980). The jets
have mass--loss rate $\gtsimeq 5\times 10^{-7}\msun~{\rm yr}^{-1}$ and
kinetic luminosity $L_k \gtsimeq 10^{39}$~erg\, s$^{-1}$
(e.g. Begelman et al., 1980; K\"onigl, 1983; Brinkmann \& Kawai, 2000;
Marshall, Canizares \& Schultz, 2002) and interact in a helical
pattern with W50 (Hjellming \& Johnston, 1981; Blundell \& Bowler,
2004).

The very regular precession of the jets is probably related to other
super--orbital periods seen in X--ray binaries. Pringle (1996, 1997)
showed that strong self--irradiation causes an accretion disc to warp
out of the orbital plane and to precess. Wijers \& Pringle (1999; see
also Ogilvie \& Dubus, 2001) showed that such discs can exhibit
precessions with the long periods seen in some X--ray binaries,
including SS433. If the disc is sufficiently warped that the mass
transfer stream does not strike the outer disc edge, then the disc
shape is such that the radiatively driven precession is retrograde
(see Figure 7 of Wijers \& Pringle, 1999).  However, such 162--day,
radiatively--driven, precessional warping occurs predominantly in the
outer parts of the disc, whereas the jets, since they are
relativistic, are presumably driven from very close to the compact
accretor at its centre (assumed to be a black hole). Thus, in this
model, it is necessary for the precession of the outer disc to
communicate itself to the centre of the disc where the initial jet
direction is determined. Indeed the problem is worse than this for two
reasons. First, if, as seems likely, the black hole is spinning, then
the inner disc is aligned with the spin of the hole (Bardeen \&
Petterson, 1975) independent of the spin of the outer disc. Second,
the precessing jets also show a nodding motion (nutation) with period
close to one--half of the 13.1~day orbital period $P_{\rm orb}$
(actually the synodic period $(2/P_{\rm orb} + 1/P_{\rm pr})^{-1}
\simeq P_{\rm orb}/2$, as the disc precession is retrograde). Disc
nodding with this period is a direct effect of the $m=2$ part of the
tidal torque (Katz et al., 1982; Bate et al., 2000) acting on the
outer part of the disc. The amplitude of such a short period
oscillation would be reduced even more severely (Katz, 1986) during
inward propagation than that of the 162--day precession, simply
because diffusion damps rapid oscillations more strongly than slower
ones. Hence there seems little hope of communicating the nodding
motion to the inner disc, where one might expect the jet directions to
be fixed (Bate et al., 2000).

In this paper we put forward a solution to this problem. We propose
that the powerful $1500$~km\, s$^{-1}$ outflow comes about because the
mass transfer rate is hyper--Eddington ($\sim 5000 \me$), and that the
outflow is therefore radiatively driven from well outside the
Schwarzschild radius, i.e.  $\ro \gg \rs$ (Shakura \& Sunyaev,
1973). (By $\me$ we mean the rate at which spherical accretion on to a
black hole with radiative efficiency $\eta \simeq 0.1$ produces the
Eddington luminosity.) The direction of the outflow is determined by
the tilt of the disc at this large radius and is subject to both the
precessional and nutational motions of the outer disc. The jets,
however, are still launched from close to the accretor with initial
direction fixed by the spin of the hole, but with final direction
aligned with the powerful (precessing and nutating) outflow.

Such a high mass transfer rate is to be expected if SS433 is a direct
descendant of a high--mass X--ray binary (HMXB) in which the more
massive companion star now fills its Roche lobe and is transferring
mass on a thermal timescale (King, Taam \& Begelman, 2000).  A
consequence of this model is that SS433 would look far brighter if
viewed along the axis of the outflow, and we suggest that in this case
it might resemble an ultraluminous X--ray source (ULX).

In Section 2, we consider the structure of the disc in SS433. We show
that at these high accretion rates, disc warp is communicated by
wave-like motions rather than by diffusion (viscous torques). In
Section 3, we discuss the nature and magnitude of the powerful
outflow, and also argue that at such high accretion rates the
bolometric luminosity of SS433 can considerably exceed $\le$. In
Section 4, we show that the outflow is massive enough to deflect the
jets and align them with the axis of the outflow. In Section 5, we
argue that SS433 would resemble a ULX, if observed along the axis of
the outflow, and suggest that hyper--Eddington mass--transfer might be
a general explanation of the ULX phenomenon. In Section 6, we suggest
that systems like SS433, but with slightly lower mass transfer rates,
might resemble the obscured {\it INTEGRAL} sources. We discuss general
implications in Section 7.

\section{The Accretion Disc in SS433}

As representative values we take the mass of the accretor to be $M_1 =
10 $ M$_\odot$, and the mass of the companion to be $M_2 = 20$
M$_\odot$. For a binary period of 13.1 days this gives a binary
separation of $a = 7.4 \times 10^{12}$ cm.  Wijers \& Pringle (1999)
show that to account for the retrograde precession, the accretion disc
in SS433 would have to be warped significantly out of the orbital
plane by radiation torques. In this situation, much of the mass
transferred by the companion joins the disc near the circularization
radius (Flannery 1975) $R_h \simeq 0.07 a = 5.2 \times 10^{11}$~cm,
rather than near its edge (Paczynski 1977) $R_{\rm tide} \simeq 0.22 a
\simeq 3R_h \simeq 1.6 \times 10^{12}$ cm. This then leads to a
central disc aligned with the binary angular momentum, and an outer
disc tilting to large angles (see Figure 7 of Wijers \& Pringle,
1999). The latter is still roughly as massive as a disc fed at its
outer edge (Wijers \& Pringle, 1999) because it acts as the reservoir
for the angular momentum lost by the matter accreting through the
inner disc, before passing this back to the companion star via tides
(as in all discs in binary systems). Thus its surface density $\Sigma$
must match on to that of the accreting inner disc at the
interface. Since there is little accretion through the outer disc, the
usual disc diffusion equation implies that $\nu\Sigma \propto
R^{-1/2}$ rather than $\nu\Sigma \simeq$~constant. Hence the total
mass and angular momentum in this outer nonaccreting disc is only
slightly less than in a normal accretion disc.

The large expected tilt angle of the outer disc may be the reason
for the otherwise puzzling deduction (Stewart et al., 1987) that the
disc partially blocks the observer's view of the central X--ray jet at
certain precession phases. Stewart et al. interpreted their result in
terms of a thick disc with aspect ratio $H/R \sim 0.4$. In order to
cause the same obscuration, a thin tilted disc would have to lie at an
angle $\tan^{-1} 0.4 \simeq 20^{\circ}$ to the orbital plane. We note
that this angle is remarkably close to the half angle $\theta$ of the
jet precession cone, suggesting that the jet somehow becomes aligned
with the axis of the tilted disc. This again brings up the question of
communicating the tilt inwards.

If the 6.3--day wobble were propagated inwards by viscosity, then it
would be damped out over a distance such that the viscous timescale
from the outer disc is of order 6.3 days. Since the dynamical
timescale at the outer disc edge is only a few days, and since the
viscous timescale is longer than the orbital timescale by a factor of
$\sim (R/H)^2 \alpha^{-1} \gg 1$, propagation of the nutation period
could not proceed very far.  However, in a nearly Keplerian disc, tilt
is communicated by bending waves, rather than viscosity, provided that
\begin{equation}
{H\over R} \ga {\rm max} (\alpha, | \Omega - \kappa |
/ \Omega),
\end{equation}
where $\alpha$ is the dimensionless viscosity parameter, $\Omega$ the
angular frequency of the disc and $\kappa$ the local epicyclic
frequency (see the discussion in Wijers \& Pringle, 1999). For the
parameters of SS433, we find, using the formulae given by Wijers \&
Pringle (1999), that at the outer disc edge $| \Omega - \kappa | /
\Omega \simeq 0.03$. In addition, for an accretion rate of $\dot{M}
\sim 10^{-5} - 10^{-3}$ M$_\odot$ year$^{-1}$, we find that in the
outer disc $H/R \simeq 0.1$ (Shakura \& Sunyaev, 1973). Thus, for
typical values of $\alpha$ in the range 0.01 -- 0.1, we see that warp
propagation can occur via bending waves in SS433.$\!$~\footnote{Since
the analysis of Wijers \& Pringle (1999) assumed that disc tilt is
communicated only through viscosity, strictly speaking it is necessary
for that analysis to be reworked for the case of SS433. We shall
assume here, however, that the basic conclusions still stand.}  How
far inwards the nutational wobble and the precessional warp can
propagate is not easy to determine. The warp waves are damped by
viscosity, and also perhaps by internal hydrodynamical instabilities
(see the discussion in Bate et al., 2000). But the wave amplitude can
also grow, by conservation of wave action, depending on the density
profile of the disc.

It is unlikely, however, that the tilt can propagate inwards as far as
regions close to the central black hole. If the hole has non-zero
spin, then the disc near the black hole must be aligned with its spin
axis out to a warp radius of a few tens of Schwarzschild radii $\rs =
3 \times 10^6$ cm (cf. King et al., 2005). If we adopt the usual
assumption that the black hole has its spin aligned with the orbital
plane,~\footnote{This appears not to be the case for some of the
microquasars, but for SS433 the accretion rate is so high that
alignment may have had time to occur (Maccarone, 2002).} then since
the inner disc is the region from which the jets must be launched, the
jets initially move along the black hole (and orbital) spin axis. We
thus arrive at a picture of the accretion disc in SS433 as tilted at
angles $\theta \simeq 20^{\circ}$ at large radii, and aligned with the
orbital plane and black hole spin close in. As we have remarked,
bending waves are able to propagate the disc tilt and its nodding
motion inwards to parts of the disc gaining mass from the companion to
some extent, but not to the innermost parts where the jets are
launched. We thus need some other agency to bend the jets to align
them with the axis of the outer disc, and thus to communicate the disc
precession and nodding to them.

\section{The Outflow from SS433}

The ultimate cause of the unique features of SS433 is probably its
extremely high mass transfer rate, which far exceeds the Eddington
rate for a $10\msun$ black hole accretor. Only a small part of the
transferred mass is lost in the jets ($\dot M_j \simeq 5\times
10^{-7}\msun\, {\rm yr}^{-1}$). Evidently most of it is expelled as
the massive high--speed outflow ($ v_w \simeq 1500$~km\, s$^{-1}$;
Fabrika 2006) seen in the form of the `stationary' H$\alpha$ line and
broad absorption lines. This velocity suggests that the outflow is
expelled from a radius (van den Heuvel, 1981; Seifina et al., 1991)
\begin{equation}
\ro \simeq {2GM\over v_w^2} \simeq 
\left({c^2\over v_w^2}\right)\rs \simeq 1.2\times 10^{11}~{\rm cm}.
\end{equation}
Support is given to these ideas by the numerical simulations of Okuda
et al. (2005).  Because accretion is highly super--Eddington, we
expect $\ro$ to correspond to the `spherization radius' given by
Shakura \& Sunyaev (1973; from their eq. 7.1) as
\begin{equation}
R_{\rm sp} = {27\mo\over 4\me}\rs
\end{equation}
(this is close to the trapping radius). Hence for this picture to be
consistent we require
\begin{equation}
\mo = {4c^2\over 27v_w^2}\me = 5000\me \simeq 5\times 10^{-4}\msun\,
{\rm yr}^{-1}.
\end{equation}
This is consistent with evolutionary calculations of the mass transfer
rate $-\dot M_2 \simeq \mo$ (King et al, 2000), and would suggest a
mass transfer lifetime $t_{\rm ev} \simeq 4\times 10^4$~yr for a
companion of mass $M_2 \simeq 20\msun$. It is also consistent with
estimates of the gas flow rate from SS433 (van den Heuvel, 1981;
Shklovskii, 1981; Blundell et al., 2001), and agrees with the estimate
of $\mo$ by King et al. (2000), which used the emission measure of the
stationary H$\alpha$ and associated infrared free--free radiation.
Things would change very little with other assumed masses (e.g. $M_1 =
2.9\msun, M_2 = 10.9\msun$, Hillwig et al., 2004, who combine
absorption--line velocities from the companion with antiphased emission--line
velocities from the vicinity of the accretor). The main change is that
the smaller accretor mass would give a smaller Eddington limit and
outflow rate (both by a factor $\sim 3$). These are still compatible
with the evolutionary calculations of King, Taam \& Begelman (2000),
and our conclusions would remain qualitatively unchanged.

\subsection{The Limiting Luminosity of SS433}

While most of the outflow from SS433 is expelled from $\ro$, we expect
further outflow from the disc inside this radius. If each disc radius
is close to its local Eddington limit, the outflow must be arranged so
that the accretion rate at disc radius $R$ decreases as $\dot M(R)
\simeq \me(R/R_{\rm in})$, where $R_{\rm in} \sim (1 - 3)\rs$ is the
innermost disc radius. The emission per unit disc face area thus goes
as $\sim 3GM\dot M(R)/8\pi R^3$. Integrating this over the disc within
$\ro$ then shows (Shakura \& Sunyaev, 1973; see the first paragraph on
p.353) that the total accretion luminosity $L_{\rm acc}$ can exceed
$\le$ by a factor $\sim \ln(\ro/R_{\rm in}) \sim \ln(-\dot M_2/\me)
\sim 10$, i.e. $L_{\rm acc}$ can be as high as $\sim 10^{40}$~erg\,
s$^{-1}$ for our assumed $10\msun$ black hole accretor. As we remark
below, the apparent accretion luminosity can exceed this if the
radiation is beamed.

\section{The Jets of SS433}

We can now understand how the precession and nodding motions are
transmitted to the jets in SS433. While bending waves cannot propagate
inward to disc radii at the base of the jets, they may well reach the
much larger radius $\ro \sim 1.2\times 10^{11}$, which is about $\sim
1/10$ of the full tilted disc radius. It is the disc orientation at $R
\sim \ro$ that determines the axis of the outflow, and thus the
outflow axis can follow both the precession and the nodding motions.
The outflow from radius $\ro$ is driven by radiation pressure
initially in a direction perpendicular to the local disc plane, and is
therefore initially collimated around the axis of the tilted disc
(cf. Okuda et al., 2005), although presumably at larger radii, $r \gg
\ro$, the outflow becomes more quasi--spherical. For the purposes of
illustration, we suppose that the outflow resembles a cylindrical
container of radius $\ro$, which is tilted with respect to the binary
(and black hole) axis by an angle of $\theta \simeq 20^\circ$.

A jet launched from very close to the black hole at the disc centre
must then collide with the outflow, at a height above the disc of $H
\simeq \ro/\tan\theta = 2.5\ro$. At this height we may assume that the
outflow is still essentially mostly flowing parallel to the tilted
disc axis. Depending on the relative ram pressures of the jet and
wind, the jet either punches its way through the container, or is
deflected to flow out parallel to its axis. To decide which, we model
the outflow `container' simply as a geometrically thin wall with
surface density
\begin{equation}
\Sigma_{\rm outflow} = {\mo\over 2\pi\ro v_w}.
\end{equation}
The jets hit this wall at angle $\theta$. We now check if this causes
significant outward deviation (i.e. along a cylindrical radius) in the
outflow. In order for the jet to be deflected sufficiently, the jet
impact must cause the flow in the wall to deflect through an angle
$\ll \sin \theta$. Thus we require the wall flow to acquire a radial
component of velocity $\Delta v \ \ll v_w \sin \theta$. The jet
momentum flux density normal to the wall is
\begin{equation}
{\rho_j v_j^2\sin\theta\over 1/\sin\theta},
\end{equation}
where $\sin\theta$ in the numerator gives the radial component (in
cylindrical coordinates), and $1/\sin\theta$ in the denominator comes
from the fact that this is spread over a larger wall area because of
the projection. An element of the outflow passing through the region
where the jet hits feels the jet's momentum flux for a time
\begin{equation}
\Delta t = {l/\sin\theta\over v_w},
\end{equation}
where $l$ is the diameter of the jet as it hits the wall. The jet collision
thus imparts sideways velocity 
\begin{equation}
\Delta v = {\rho_j v_j^2\sin^2\theta\over \Sigma_{\rm
    outflow}}.{l/\sin\theta\over v_w}.
\end{equation}
If the {\it full} opening angle of the jet is $\theta_j$ we have
$l=R\sin\theta_j$, where $R = \ro/\sin\theta$. Mass conservation
requires that
\begin{equation}
\dot M_j = {\pi\over 2}\rho_jv_jl^2,
\end{equation}
where we have assumed that $\frac{1}{2} \dot M_j$ is the mass flux in
each jet.
We thus find
\begin{equation}
{\Delta v \over v_w} = {2\over \pi}{\dot M_jv_j\over l^2}{l\sin\theta\over
  v_w^2}{2\pi\ro v_w\over \mo}
\end{equation}
which reduces to 
\begin{equation}
{\Delta v\over v_w} = 4{\dot M_jv_j\over \mo
v_w}{\sin^2\theta\over\sin\theta_j}.
\end{equation}
With $\dot M_j \sim \me, v_j \sim c$ this is
\begin{equation}
{\Delta v\over v_w} \simeq
4\left({\rs\over\ro}\right)^{1/2}{\sin^2\theta\over \sin\theta_j}.
\end{equation}
With $(\ro/\rs)^{1/2} = 200, \theta \simeq 20^{\circ}, \theta_j =
3^{\circ}$ (Fabrika, 2006) gives $\Delta v/v_w \simeq 0.045 \ll \sin \theta
\simeq 0.34$.

We conclude that the outflow is well capable of bending the jet
direction until it is parallel to the outflow axis, and thus that the
outflow can communicate the precession and nodding motions to the jet
direction. If the jet deflection takes place through one oblique
shock, then in this simple picture the flow is deflected through an
angle $\chi = \theta = 20^{\circ}$ and the post-shock velocity is the
jet velocity of $v_j \simeq 0.26c$. We can then make use of (the
special--relativistic generalization of) the standard shock strophoid
(cf. Landau \& Lifschitz, 1959, eq. 86.5) to compute the properties of
the shock.  The main result of this, assuming that the shock is very
oblique so that the post-shock velocity is supersonic, is that only
about $\sim 1$\% of the preshock kinetic energy, i.e.  $\sim
10^{37}$~erg\, s$^{-1}$, is dissipated in the shock.$\!$~\footnote{This
energy would be recovered as the post--shock jet fluid reaccelerates
unless the shocks are sufficiently radiative -- which is likely given
the high post--shock temperatures.} It is more realistic to suppose
that the jets are deflected through a series of oblique shocks. In
this case, the amount of dissipation of jet energy is lower, and it is
now distributed along the jet. Thus the amount of energy dissipated in
deflecting the jets makes little contribution to the overall energy
output of the system, and does not slow the jets significantly. These
bending shocks, and concomitant radiative losses, are hidden from the
observer within the outflow (see below). The observed jet X--ray
emission (Watson et al., 1986, Stewart et al., 1987) comes from
further out, presumably from internal shocks, and is observable since
the outflow has spread geometrically at such distances. The only way
in which the energy dissipated in the jet-deflecting shocks might be
observable is if the post-shock temperatures ($\sim 10^{11}$ K) give
rise to high energy (e.g. MeV) emission, and if the system were viewed
down the axis of the outflow.

Evidently the preshock jet is much slower than those with Lorentz
factors $\gamma \ga 10$ inferred in sub--Eddington, or near-Eddington
($\dot{M} \ltsimeq \me$), accreting systems (Miller-Jones, Fender \&
Nakar, 2006). It is not clear why this is so, but it may indicate that
because in SS433 the accretion is strongly super-Eddington ($\dot{M}
\gg \me$), the jets in this case are accelerated by radiation
pressure, rather than by magnetic processes (cf. the outflows computed
by Okuda et al., 2005).

Our picture makes a further prediction. We would not expect the tilt
angle $\theta$ of the outer disc and outflow to be precisely constant
in time, and indeed small variations are seen. If the bending shocks
are sufficiently radiative --- which seems likely given the high
post--shock temperatures --- the jets conserve only the velocity
component $\propto \cos\theta$ parallel to the outflow axis. We would
therefore expect a cosinusoidal anticorrelation, i.e.
\begin{equation}
v_j \propto 1/ \cos \theta,
\end{equation}
between the observed jet velocity and the precession cone angle
$\theta$. Such an anticorrelation is seen (Blundell \& Bowler, 2005).

\section{SS433 as a ULX}

We can now see why SS433 appears so underluminous compared with its
mechanical energy output. Most of its radiative luminosity must be
produced within $\ro$ of the accretor, i.e. within the outflow. The
Thomson optical depth radially across this outflow is $\tau_{\perp} =
\kappa_{es}\Sigma_{\rm outflow} \simeq 110$, whereas its central
regions are essentially transparent along the flow axis,
i.e. $\tau_{\parallel} \la 1$. Most of the luminosity must therefore
escape in directions close to this axis, collimated into a solid angle
$\Omega$ set by the geometric spreading of the outflow away from the
axis. An observer within this cone would measure a radiation flux
density given by $\sim 10\le/(4\pi/\Omega)$ rather than $\le$ and thus
identify the source as ultraluminous.

This suggests that one basic cause of ULX behaviour is collimation of
the outgoing radiation (King et al., 2001), and that, as suspected for
some time (King et al., 2001; King, 2002; Charles et al., 2004), SS433
is a ULX viewed `from the side'. The total luminosity escaping
sideways (i.e. from the outside of the `container') is $\le$, but
thermalized over an area $\sim 4\pi \ro^2$. This gives an effective
temperature $T_{\rm eff} \sim 10^5$~K. While this radiation is enough
to ensure that the outer disc is efficiently warped, it is far too
soft to be observable through the interstellar photoelectric
absorption column ($7\times 10^{21}~{\rm cm}^{-2}$, Kotani et al.,
1996) towards SS433.  The picture of ULXs presented here differs from
the scaled--down BL Lac object (K\"{o}rding et al., 2002) sometimes
invoked, in that, in our picture the radiation is collimated by
scattering, rather than relativistically beamed. Moreover the jets
produced near the accretor have to make their way out through the
outflow, and also appear to be definitely sub--relativistic.

We note that some ULXs are surrounded by nebulae which are too large to be
supernova remnants and appear to be collisionally ionized (Gris\'e et al.,
2006a, b; Pakull et al., 2006). These are reminiscent of the W50 nebula around
SS433, reinforcing our suggestion that ULXs are super--Eddington accreting
binaries like SS433.

It is not clear how to estimate {\it a priori} the solid angle
$\Omega$, the inverse of which specifies just how ultraluminous a ULX
can be, without undertaking a numerical simulation involving radiation
hydrodynamics (cf. Okuda et al., 2005).  Below we argue that to
achieve consistency between the number of progenitor HMXBs in our
Galaxy ($\sim 10$) and the number of ULXs ($\la 1$) we require
$\Omega/4 \pi \la 1/10$ for the latter.  This would give a peak
apparent luminosity for a hyper-accreting $10 \msun$ black hole of
$\ga 10^{41}$ erg s$^{-1}$.

The radiation spectrum predicted for ULXs by this picture is a
combination of the usual complex medium--energy X--ray binary
spectrum, plus a very strong soft component resulting from
thermalization of a large fraction of the accretion luminosity $\sim
10 \le$ over the walls of the outflow, i.e. over an area $\sim
\pi\ro^2$, but beamed into a solid angle $\Omega$. This gives a beamed
flux with colour temperature $T_c \sim 3 \times 10^5$ K, assuming
$\Omega/4\pi \sim 0.1$.  Strong soft components of this type are
indeed seen in ULXs with sufficiently low absorption columns
(e.g. Miller et al., 2003). In our picture, the large blackbody radius
$R_b$ deduced from this component gives an estimate of the
super--Eddington factor as
\begin{equation}
{\mo\over\me} \simeq {4R_b\over 27\rs}, 
\end{equation}
rather than an estimate of the black hole mass from the assumption
that $R_b \simeq 3\rs$. Thus the IMBH masses $\ga 10^3\msun$ deduced
by Miller et al (2003) for the ULXs NGC 1313 X--1, X--2 become in our
picture estimates of the super--Eddington factor $\mo/\me \ga 450$ for
a $10\msun$ black hole.

Strohmayer \& Mushotzsky (2003) use the occasional presence of a 54
mHz (period $\sim 18 s$) QPO seen in the 2 -- 10 keV flux of the ULX
M82 X--1 as evidence against this source being highly beamed, and in
favour of it being an intermediate mass black hole (IMBH). There are
two main points to the argument. First, the high rms amplitude of the
QPO ($\sim 8$ per cent) would be washed out if the beaming process
involved a lot of scattering.  In our picture of SS433 most of the
luminosity emerges over a collimating region of size $\sim 4$ light
seconds. The fact that the QPO in M82 X--1 is observed in the 2 -- 10
keV energy range suggests that is it observed in photons which come
from close to the black hole and which have not undergone a large
amount of scattering. Thus we are most likely looking down the central
regions of the collimating outflow, which are essentially transparent
($\tau_{\parallel} \la 1$, see above). These regions would not
significantly wash out a QPO with period $\sim 18$s unless there was
significant optical depth (Kylafis \& Klimis, 1987). Second, if one
assumes that the QPO frequency scales with mass of the compact object,
then comparision with typical QPO frequencies of $\sim 0.8 - 3$ Hz in
stellar mass objects suggests a mass of $\sim 100 - 300$ M$_\odot$
for this object. Strohmayer \& Mushotzsky admit that the second point
is a weak one and that `the broadband variability in the M82 ULX and
the lack of a nonthermal component argue against this
identification'. We note that in any case there is as yet no
believable physical picture of how QPOs are made, and also that,
because of the lack of flux, detecting a QPO at around 1Hz in M82 X--1
would be problematic.


\section{The Obscured {\it INTEGRAL} Sources}

In SS433 the walls of the outflow `container' are optically thick
($\tau_{\perp} \gg 1$). Thus most of the X--ray radiative output
emitted from the outside of this flow towards a terrestrial observer
of SS433 is degraded to photon energies too low to be observable
through the ISM. This may help explain the difficulty in finding
similar `sideways--on' systems as X--ray sources, even though there
should be roughly as many of these as there are HMXBs. However, if the
mass transfer rate were rather less super--Eddington than in SS433,
the total scattering optical depth $\tau_{\perp}$ across the outflow
would be lower. For $\tau_{\perp} \la 7$ the combined effects of
scattering and absorption do not completely thermalize the emerging
radiation, which might thus appear as a very hard X--ray continuum
with a high intrinsic absorption column $N_H \la {\rm few}\times
10^{24}$~cm$^{-2}$, presumably accompanied by powerful fluorescent
emission lines. This is just what is seen in the new class of obscured
HMXB systems found by {\it INTEGRAL} (Revnivtsev et al., 2003; Walter
et al., 2003; Dean et al., 2005). Note that our picture of
hyper--Eddington accreting sources applies also to neutron--star
systems, as many of the {\it INTEGRAL} sources appear to be, with
luminosity limits lower by the ratio of neutron--star to black--hole
mass. Some of these obscured sources may be modulated (either in
scattering or absorption) at one-half of the orbital period because of
the $m=2$ tide from the companion star (cf. the 6.3--d nodding motion of
the jets in SS433).

\section{Discussion}

We have shown that the effects of the powerful outflow expelling the
super--Eddington mass transfer in SS433 can explain several of aspects
of this system. The outflow is able to bend the jets from the accretor
and redirect them along the axis of the outer disc, which is tilted by
radiation torques. As a result the jets show both the 162~d precession
and the 6.3~d nodding motion driven by the companion star's tide. Our
picture also explains the observed anticorrelation of precession cone
angle and jet velocity (Blundell \& Bowler, 2005).

This picture provides an explanation of why SS433 is faint in observed
electromagnetic radiation (especially X--rays) compared with the
kinetic luminosity of the jets ($L_k \gtsimeq 10^{39}$ erg s$^{-1}$).
Indeed, we have shown that SS433 gives a plausible picture of ULX
behaviour if we imagine viewing it along the outflow axis.  The
radiative luminosity exceeds the formal Eddington value $\le$ by a
factor $\sim 10$, and most of this luminosity is radiated in a cone
around the outflow axis. This picture can account for apparent ULX
luminosities $\gg \le$ from stellar--mass binaries. High--mass X--ray
binary systems undergoing thermal--timescale mass transfer like SS433
are inevitable in galaxies with vigorous star formation, accounting
for the observed correlation of ULXs with star--forming regions in
such galaxies, and their birthrate is known to be of the right order
to explain the incidence of ULXs (King et al., 2001), independently of
the collimation solid angle $\Omega$. Very bright outbursting soft
X--ray transients may reproduce some of these features, as the
accretion rates driven by disc instabilities can be highly
super--Eddington for some time (cf. Cornellisse, Charles \& Robertson,
2006). These may produce transient ULXs in early--type galaxies (King
2002).

In addition, this picture helps to explain the uniqueness of SS433
within our own galaxy. We can get some idea of the collimation solid
angle $\Omega$ by considering the progenitors of systems like
SS433. These are HMXBs just beginning Roche--lobe overflow
(e.g. Verbunt \& van den Heuvel, 1995). After $\sim 10^5$~yr the mass
transfer rate becomes super--Eddington and the system becomes an
SS433--like object. Other HMXBs such as the Be--X--ray binaries and
supergiant wind--fed systems appear less likely to evolve into systems
like SS433,$\!$~\footnote{The Be--X--ray binaries are generally too
  wide for full Roche lobe overflow to occur, and the supergiant
  systems are likely to merge.}
so we estimate the current number of SS433 progenitors in
the Galaxy as $\sim 10$. The thermal-timescale mass--transfer stage
characterizing SS433 should last at least as long as the preceding
HMXB stage, and should exhibit a comparable accretion luminosity. So,
were SS433--like systems to emit isotropically, we might have expected
to find as many SS433--like systems as relevant HMXBs (i.e. $\sim 10$)
in the Galaxy. The fact the we have only one SS433--like object, which
we suggest is a ULX seen from the side, probably tells us that such
objects are hard to detect.  However, the fact that we see no ULXs,
which we identify as SS433--like objects collimated towards us and
which should be easily detectable X--ray sources, suggests that the
mean solid angle into which most of the radiation is emitted is
$\Omega / 4\pi \la 1/10$. It may be possible to get better estimates
of the mean value of $\Omega$ by comparing the numbers of ULXs and
HMXBs in galaxies where both are detected. We note that it is to be
expected that $\Omega$ varies among ULXs, possibly as a function of
the Eddington ratio.

As a further check on our adopted value of $\mo$, we note that the
outflow is evidently quasi--spherical at large distances from the
binary, and drives a wind bubble into the interstellar medium. It is
easy to check that this must be in the energy--driven phase, giving a
radius
\begin{equation}
R_{\rm bubble} \simeq 0.9\left({\mo v_w^2\over
2\rho}\right)^{1/5}t^{3/5} = 33\rho_{26}^{-1/5}\left({t\over
2t_{\rm ev}}\right)^{3/5}~{\rm pc}
\end{equation}
where $\rho = 10^{-26}\rho_{26}~{\rm g\, cm^{-3}}$ is the mass density
of the ISM. We see that the outflow may well have inflated the almost
perfectly spherical `head' of the W50 nebula (whose observed radius is
$\simeq 42$~pc, see Fabrika 2006 and references therein) within its
mass transfer lifetime. (Note that the precessing jets inflate the
`ears' of W50 still more.)

The jet precession, together with the tidally induced nutation, is a
useful diagnostic in SS433. However, it is important to realise that
regular precession is not generic to the ULX picture we have set out
(Wijers \& Pringle, 1999; Ogilvie \& Dubus, 2001). For example, in
other similar objects, and in SS433 itself at other epochs, the black
hole spin may not be aligned with the binary orbit, the disc may not
be tilted by radiation torques, and even if tilted may not precess,
let alone regularly. If we were lucky enough to lie in the ULX
visibility cone of a system undergoing regular precession like SS433,
this would appear as a ULX with a regular period of several tens of
days, but with a rather short duty cycle and consequent low discovery
probability. If, instead, an observed ULX has irregular disc
precession, as is more usual in Galactic HMXBs, then this would
translate into irregular X--ray variability, including abrupt
disappearances and reappearances. If the precession angle is small (or
zero), the X--rays would be on all the time. M82 X--1 may be of this
type.  If confirmed, the 62--d timescale recently claimed for it
(Kaaret, Simet \& Lang, 2006) seems more likely to be a precession
quasi--period than the orbital period suggested (note that there are
at least 6 X--ray binaries with super--orbital periods shorter than
the claimed one, and 8 less than 162~d (cf. Wijers \& Pringle,
1999)). We reiterate that the spin axis of the accretor need not be
aligned with the binary, particularly if the accretor was formed in a
supernova explosion (see Maccarone, 2002; King et al., 2005 for
discussion of black--hole alignment). This may be another cause of
irregular X--ray variability in ULXs. Given all these possibilities,
long--term X--ray monitoring of ULXs appears worthwhile.

Our picture suggests that some ULXs might show mildly blueshifted
spectral features. Velocities such as the $0.26c$ in SS433 are
probably at the limit of current observational capabilities, but would
offer major insight if detected. Similarly, it is probably optimistic
to hope to see 0.5 MeV pair emission from a jet bending shock in ULXs
(the oblique shock produces near--relativistic ion temperatures $\sim
10^{11}$~K), particularly if this proves undetectable in edge--on
Galactic systems.

\section{Acknowledgments} 

MCB acknowledges support from NSF grant AST-0307502 and from the
University of Colorado Council on Research and Creative Work. He
thanks the Institute of Astronomy and Trinity College, Cambridge for
hospitality.  ARK gratefully acknowledges a Royal Society Wolfson
Research Merit Award. He thanks Hans Ritter for stimulating
conversations on several aspects of this work over a number of years.


\begin{thebibliography}{}

\bibitem{} Abell G.O.,  Margon B., 1979, Nature, 279, 701

\bibitem{} Bardeen J.M., Petterson, J.A., 1975, ApJ, 195, L65

\bibitem{} Bate M.R., Bonnell I.A., Clarke C.J., Lubow S.H., Ogilvie
G.I., Pringle J.E., Tout C.A., 2000, MNRAS, 317, 773

\bibitem{} Begelman M.C., Sarazin C.L., Hatchett S.P., McKee C.F.,
Arons J., 1980, ApJ, 238, 722

\bibitem{} Blundell K.M, Mioduszewski A.J., Podsiadlowski P., Muxlow
T.W.B., Rupen M.P., 2001, ApJ, 562, L79

\bibitem{} Blundell K.M., Bowler M.G., 2004, ApJ, 616, L159

\bibitem{} Blundell K.M., Bowler M.G., 2005, ApJ, 622, L129

\bibitem{} Brinkmann W., Kawai N., 2000, A\&A, 363, 640

\bibitem{} Charles, P.A., Barnes, A.D., Casares, J., Clark, J.S.,
Clarkson, W.I., Harlaftis, E.T., Hynes, R.I., Marsh, T.R., Steeghs,
D., 2004, in Compact Binaries in the Galaxy and Beyond, Edited by
G. Tovmassian and E. Sion. Revista Mexicana de Astronomía y
Astrofísica (Serie de Conferencias) Vol. 20. IAU Colloquium 194,
pp. 50-52 

\bibitem{} Cornelisse R., Charles P.A., Robertson C., 2006, MNRAS,
366, 918

\bibitem{} Dean, A.J., Bazzano, A., Hill, A.B., Stephen, J.B.,
Bassani, L., Barlow, E.J., Bird, A.J., Lebrun, F., Sguera, V., Shaw,
S.E., Ubertini, P., Walter, R., Willis, D.R., 2005, A\&A 443, 485

\bibitem{} Fabian, A.C., Rees M.J., 1979, MNRAS, 187, 13P

\bibitem{} Fabrika S., 2006, astro-ph/0603390

\bibitem{} Flannery B.P., 1975, MNRAS, 170, 325

\bibitem{} Gris\'{e} F., Pakull M.W., Motch C., 2006a,  in Proc. IAU
Symp 230, `Populations of High Energy sources in Galaxies', eds. E.J A
Meurs, G. Fabbiano, astro--ph/0603768

\bibitem{} Gris\'{e} F., Pakull M.W., Motch C., 2006b, in 'The X-ray
Universe 2005',  astro--ph/0603769

\bibitem{} Hillwig, T.C., Gies, D.R., Huang, W., McSwain, M.V., Stark,
  M.A., van der Meer, A., Kaper, L., 2004, ApJ 615, 422

\bibitem{} Hjellming R.M., Johnston K.J., 1981, Nature, 290, 100

\bibitem{} Kaaret P., Simet M.G., Lang C.C., 2006, ApJ, in press
(astro-ph/0604029) 

\bibitem{} Katz J.I., 1986, Comments on Ap, 11, 201

\bibitem{} Katz J.I., Anderson S.F., Margon B., Grandi S.A., 1982,
ApJ, 260, 780

\bibitem{} King A.R., 2002, MNRAS, 335, L13

\bibitem{} King A.R., Davies M.B., Ward M.J., Fabbiano G., Elvis M.,
2001, ApJ, 552, L109

\bibitem{} King A.R., Lubow S.H., Ogilvie G.I., Pringle J.E., 2005,
MNRAS, 363, 49

\bibitem{} King A.R., Taam R.E, Begelman M.C, 2000, ApJ, 530, L25

\bibitem{} K\"{o}rding E., Falcke H., Markoff, S., 2002, A\&A, 382, L13

\bibitem{} K\"onigl, A., 1983, MNRAS, 205, 471

\bibitem{} Kotani T., Kawai, N., Matsuoka, M.,Brinkmann, W., 1996,
PASJ, 48, 619

\bibitem{} Kylafis, N.D., Klimis, G.S., 1987, ApJ 323, 678

\bibitem{} Landau L.D., Lifschitz E.M, 1959, Fluid Mechanics, Pergamon
  Press, Oxford

\bibitem{} Maccarone T., 2002, MNRAS, 336, 1371

\bibitem{} Mammano A., Vittone A., 1978, IAU Circ. N3314, 2

\bibitem{} Margon B., 1979, IAU Circ. N3345, 1

\bibitem{} Margon B., 1984, ARA\&A, 22, 507

\bibitem{} Marshall H.L., Canizares C.R., Schultz N.S., 2002, ApJ,
564, 941

\bibitem{} Milgrom, M., 1979, A\&A, 78, L9

\bibitem{} Miller J.M., Fabbiano G., Miller M.C., Fabian A.C.,
  2003, ApJ 585, L37

\bibitem{} Miller-Jones J.C.A., Fender R.P., Nakar E., 2006, MNRAS,
  367, 1432


\bibitem{} Okuda T., Teresi V., Toscano E., Molteni D., 2005, MNRAS,
357, 295

\bibitem{} Ogilvie G.I., Dubus G., 2001, MNRAS, 320, 485


\bibitem{} Paczynski B., 1977, ApJ, 216, 822 

\bibitem{} Pakull M.W., Gris\'{e} F., Motch C., 2006, in `The X-ray
Universe 2005', astro--ph/0603771

\bibitem{} Pringle J.E., 1996, MNRAS, 281, 357

\bibitem{} Pringle J.E., 1997, MNRAS, 292, 136

\bibitem{} Revnivtsev M.G., Sazunov, S.Y., Gilfanov M.R., Sunyaev
R.A., 2003, Astronomy Letters, 29, 587

\bibitem{} Seifina E.V., Shakura N.I., Postnov K.A., Prokhorov M.E.,
1991, Lect. Notes Phys., 385, 151

\bibitem{} Shakura N.I., Sunyaev R.A., 1973, A\&A, 24, 337


\bibitem{} Shklovskii I.S., 1981, Sov Astron, 25, 315

\bibitem{} Stewart G.C., Watson M.G., Matsuoka M., Brinkmann W.,
Jugaku J., Takagishi K., Omodaka T., Kemp J.C., Kenson G.D., Kraus
D.J., Mazeh T., Leibowitz E.M., 1987, MNRAS, 228, 293 

\bibitem{} Strohmayer T., Mushotzky R.F., 2003, ApJ, 586, L61

\bibitem{} Verbunt, F., vand den Heuvel, E.P.J., 1995, chapter 11 of
X--ray Binaries, Cambridge University Press

\bibitem{} van den Heuvel, E.P.J., 1981, Vistas Astron, 25, 95

\bibitem{} Watson M.G., Stewart G.C., King A.R., Brinkmann W.,  1986,
MNRAS, 222, 261

\bibitem{} Walter R., Rodriguez J., Foschini L., de Plaa J., Corbel
S., Courvoisier T.J-L., den Hartog P R., Lebrun F., Parmar A.N.,
Tomsick, J.A., Ubertini P., 2003, A\&A, 411, L42

\bibitem{} Wijers R.A.M.J., Pringle J.E., 1999, MNRAS, 308, 207

\end{thebibliography}
\end{document}